\documentclass[preprint,showpacs,showkeys,preprintnumbers,amsmath,amssymb,aps]{revtex4-1}
\usepackage{bm,amssymb,amsmath,mathrsfs}
\usepackage[usenames]{color}
\usepackage{bookmark}

\usepackage{dcolumn}% Align table columns on decimal point

\newcommand{\bc}{\begin{center}}
\newcommand{\ec}{\end{center}}
\newcommand{\bit}{\begin{itemize}}
\newcommand{\eit}{\end{itemize}}
\newcommand{\bq}{\begin{equation}}
\newcommand{\eq}{\end{equation}}

\begin{document}

\title{Relativistic Stern-Gerlach Deflection: Hamiltonian Formulation}

\author{S.~R.~Mane}
\email{srmane001@gmail.com}

\affiliation{Convergent Computing Inc., P.~O.~Box 561, Shoreham, NY 11786, USA}

\begin{abstract}
A Hamiltonian formalism is employed to elucidate the effects of the 
Stern-Gerlach force on beams of relativistic spin-polarized particles,
for passage through a localized region with a static magnetic or electric field gradient.
The problem of the spin-orbit coupling for nonrelativistic bounded motion in a central potential 
(hydrogen-like atoms, in particular) is also briefly studied.
\end{abstract}



\maketitle

\setcounter{equation}{0}
\section{\label{sec:intro} Introduction}
The Stern-Gerlach force for a beam of spin-polarized particles is usually treated nonrelativistically,
e.g.~in the classic text by Mott and Massey \cite{MottMassey}.
See also the text by Kessler \cite{Kessler}, 
who essentially reproduces the analysis in \cite{MottMassey} 
and adds some commentary based on more modern theory and technology.
Nevertheless, the corresponding problem for a beam of relativistic particles, though less well studied, is also of interest.
It has long been recognized that, when the effects of special relativity are accounted for
(including systems such as the electronic orbitals in atoms) 
the spin-orbit coupling includes contributions from both a magnetic dipole interaction and Thomas precession \cite{Thomas}.
See, e.g.~the text by Jackson \cite{JacksonCE3}.
Recently, the effect of the Stern-Gerlach force on a beam of relativistic spin-polarized particles 
has been analyzed theoretically \cite{Talman_arxiv_Nov2016}.
The formulas derived in \cite{Talman_arxiv_Nov2016} indicate that the transverse momentum transfer 
imparted to a beam of particles, via the Stern-Gerlach force,
upon passage through a localized region with a static magnetic or electric field gradient, 
is proportional exclusively to the magnetic dipole moment of the particles, 
with no contribution from the Thomas precession.
The analysis in \cite{Talman_arxiv_Nov2016} also treats, via perturbation theory,
the problem of the spin-orbit coupling for bounded nonrelativistic motion in a central potential.
A nonzero energy shift is obtained for the ground state of the hydrogen atom,
where the ground state is split into a doublet according to the orientation of the electron spin
\cite[eq.~(58)]{Talman_arxiv_Nov2016}.

To aid in clarifying the above issues,
we present a derivation of the spin-dependent transverse momentum transfer 
for a beam of relativistic spin-polarized particles,
upon passage through a localized region with a static magnetic or electric field gradient.
The effects of the spin-orbit coupling for bounded nonrelativistic motion in a central potential is also briefly studied.
A Hamiltonian formalism is employed throughout.

%\vfill\pagebreak
\setcounter{equation}{0}
\section{\label{sec:hamsg} Hamiltonian and Equations of Motion}
We treat a particle of mass $m$ and charge $e$, with velocity $\bm{v}=\bm{\beta}c$ and Lorentz factor $\gamma = 1/\sqrt{1-\beta^2}$.
We treat only particles of spin $\frac12$.
The (rest-frame) spin operator is denoted by $\bm{s}$, with magnetic moment anomaly $a=\frac12(g-2)$.
The canonical coordinates and conjugate momenta are denoted by $\bm{r}$ and $\bm{p}$, respectively.
The orbital motion is treated semiclassically, but the spin is a quantized operator.
The prescribed external electric and magnetic fields, and the vector and scalar potentials,
are denoted by $\bm{E}$, $\bm{B}$, $\bm{A}$ and $\Phi$, respectively.
Radiation by the particles is neglected.
The Hamiltonian is
\bq
\label{eq:hamsg}
H = H_{\rm orb}(\bm{r},\bm{p}) + \bm{\Omega}(\bm{r},\bm{p})\cdot\bm{s} \,.
\eq
Here $\bm{\Omega}$ is the spin precession vector.
Both $H_{\rm orb}$ and $\bm{\Omega}$ depend only on $\bm{r}$ and $\bm{p}$ and explicit expressions for them will be displayed below.
The fundamental point here is that the use of a Hamiltonian determines {\em both} the spin and orbital equations of motion uniquely.
The equations of motion are, for $j=1,2,3$,
\begin{subequations}
\label{eq:eom}
\begin{align}
\label{eq:eom_r}
\frac{dr_j}{dt} &= \phantom{-}\frac{\partial H_{\rm orb}}{\partial p_j} 
+\frac{\partial (\bm{\Omega}\cdot\bm{s})}{\partial p_j} \,,
\\
\label{eq:eom_p}
\frac{dp_j}{dt} &= -\frac{\partial H_{\rm orb}}{\partial r_j} 
-\frac{\partial (\bm{\Omega}\cdot\bm{s})}{\partial r_j} \,,
\\
\label{eq:eom_s}
\frac{d\bm{s}}{dt} &= \phantom{-}\bm{\Omega} \times\bm{s} \,.
\end{align}
\end{subequations}
The spin-orbit coupling term $\bm{\Omega}\cdot\bm{s}$ 
determines {\em both} the spin precession equation of motion eq.~\eqref{eq:eom_s} 
{\em and} the spin-dependent effects on the orbital motion (the last terms in eqs.~\eqref{eq:eom_r} and \eqref{eq:eom_p}).
In particular, the last term in eq.~\eqref{eq:eom_p} is the Stern-Gerlach force.
The explicit expressions for $H_{\rm orb}$ and $\bm{\Omega}$ are
\begin{subequations}
\label{eq:horb_spin}
\begin{align}
H_{\rm orb} &= \biggl[\, \biggl(\bm{p}-\frac{e}{c}\bm{A}\biggr)^2 c^2 + m^2c^4\,\biggr]^{1/2} +e\Phi \,,
\\
\bm{\Omega} &= -\frac{e}{mc}\,\biggl[\, \biggl(a+\frac{1}{\gamma}\biggr)\bm{B}
-\frac{a\gamma}{\gamma+1}(\bm{\beta}\cdot\bm{B})\bm{\beta}
-\biggl(a+\frac{1}{\gamma+1}\biggr)\bm{\beta}\times\bm{E}\,\biggr] \,.
\end{align}
\end{subequations}
Note that $\bm{\Omega}$ is {\em not} proportional to the magnetic dipole moment if $\bm{\beta} \ne 0$. 

We close this section with the following remarks.
To make contact with the notation in \cite{Talman_arxiv_Nov2016},
we define $\hat{s}^*_j$ via $s_j = \frac12\hbar \hat{s}^*_j$ for $j=1,2,3$.
The analysis in \cite{Talman_arxiv_Nov2016} treats only electrons.
Let us define $a_e=\frac12(g-2)$ for an electron.
Then the electron magnetic dipole moment is given by \cite[eq.~(1)]{Talman_arxiv_Nov2016}
\bq
\mu_e^* = (a_e+1)\frac{e\hbar}{2m_ec} \,.
\eq
Note that we employ Gaussian units, hence the factor of $c$ in the denominator,
which does not appear in \cite[eq.~(1)]{Talman_arxiv_Nov2016}.
Curiously, in the text after defining the electron magnetic dipole moment,
it is stated in \cite{Talman_arxiv_Nov2016} that
``Furthermore there will be no discussion of subtleties such as anomalous magnetic moments.''
However, the expression for $\mu_e^*$ in \cite[eq.~(1)]{Talman_arxiv_Nov2016}
{\em does} include the electron anomalous magnetic dipole moment.

%\vfill\pagebreak
\setcounter{equation}{0}
\section{\label{sec:cov} Covariant Equations of Motion}
Dam and Ruijgrok \cite{DamRuijgrok1980}
have derived covariant classical relativistic equations of motion
for particles with spin moving in external fields.
We display their equations below, for the orbit and spin.
Following standard practice, Greek indices run from 0 to 3
and Roman indices run from 1 to 3.
They set the speed of light to unity $c=1$.
As above, the particle rest mass is $m$, the charge is $e$ and $g$ denotes the spin $g$-factor.
They employed the metric
$g_{\mu\nu} = \textrm{diag}(-1,1,1,1)$.
The particle velocity and spin four-vectors
are denoted by $u^\mu$ and $W^\mu$, respectively.
The electromagnetic field tensor is denoted by
$F^{\mu\nu}$ and its dual is
$\tilde{F}^{\mu\nu} = \frac12\epsilon^{\mu\nu\lambda\sigma}F_{\lambda\sigma}$, 
where
\bq
F^{0k} = E_k \,,\qquad
F^{kl} = \epsilon_{klm} B_m \,,\qquad
\tilde{F}^{0k} = B_k \,,\qquad
\tilde{F}^{kl} = -\epsilon_{klm} E_m \,.
\eq
A dot denotes a derivative with respect to the proper time.
The equation of motion for the four-velocity $u^\mu$ is
\cite[eq.~(3.22)]{DamRuijgrok1980}
\bq
\begin{split}
\label{eq:damruijgrok_u}
\dot{u}^\mu &= \frac{e}{m}\,F^{\mu\nu}u_\nu
-\frac{eg}{2m^2}(g^{\mu\nu} +u^\mu u^\nu) \partial_\nu \tilde{F}^{\rho\lambda} W_\rho u_\lambda
\\
&\quad
+\frac{e}{2m^2}(g-2)(g^{\mu\nu} +u^\mu u^\nu) u^\alpha\partial_\alpha \tilde{F}_{\nu\beta} W^\beta \,.
\end{split}
\eq
The equation of motion for the spin four-vector $W^\mu$ is
\cite[eq.~(3.23)]{DamRuijgrok1980}
\bq
\label{eq:damruijgrok_W}
\dot{W}^\mu = \frac{eg}{2m}\,F^{\mu\nu}W_\nu
+\frac{e}{2m}(g-2)(u^\alpha F_{\alpha\beta} W^\beta)u^\mu
-\frac{eg}{2m^2}(W^\alpha \partial_\alpha \tilde{F}_{\beta\gamma}W^\beta u^\gamma) u^\mu \,.
\eq
It was noted in \cite{DamRuijgrok1980}
that both of the above equations are consistent with the 
covariant constraints 
$u^\mu u_\mu = -1$, 
$W^\mu W_\mu$ is constant
and $u^\mu W_\mu = 0$.
\bit
\item
As noted in \cite{DamRuijgrok1980},
if the terms which contain derivatives of $F_{\mu\nu}$ are neglected,
we obtain 
\begin{subequations}
\begin{align}
\dot{u}^\mu &= \frac{e}{m}\,F^{\mu\nu}u_\nu \,,
\\
\dot{W}^\mu &= \frac{eg}{2m}\,F^{\mu\nu}W_\nu
+\frac{e}{2m}(g-2)(u^\alpha F_{\alpha\beta} W^\beta)u^\mu \,.
\end{align}
\end{subequations}
The equation for $u^\nu$ is the Lorentz force law 
and the equation for $W^\nu$ is the 
spin precession equation derived by Bargmann, Michel and Telegdi \cite{BMT},
which is known to the equivalent to that derived by Thomas \cite{Thomas}.
See the text by Jackson \cite{JacksonCE3}.

\item
The terms in eq.~\eqref{eq:damruijgrok_u}
which contain the spin $W^\mu$ and field gradients
(i.e.~derivatives of $F_{\mu\nu}$) 
constitute the `covariant Stern-Gerlach force.'
Because of the presence of the term in $g-2$,
the Stern-Gerlach force is {\em not} proportional to the particle's magnetic dipole moment.

\item
Dam and Ruijgrok noted that 
in the particle's instantaneous rest frame (irf),
the nonzero components of the term in $g-2$ in eq.~\eqref{eq:damruijgrok_u}
have the following value 
\cite[eq.~(3.24)]{DamRuijgrok1980}
($\bm{S}$ is the spin three-vector and $t$ is the time in the instantaneous rest frame)
\bq
\biggl[\,\frac{e}{2m^2}(g-2)(g^{\mu\nu} +u^\mu u^\nu) u^\alpha\partial_\alpha \tilde{F}_{\nu\beta} W^\beta \,\biggr]_{\rm irf}
= -\frac{e}{2m^2}(g-2) \bm{S} \times \frac{\partial\bm{E}_{\rm irf}}{\partial t} \,.
\eq
The above term does not appear in the analysis in \cite{Talman_arxiv_Nov2016}.
  
\item 
It is theoretically possible for a particle with spin to have a magnetic moment of zero.
Setting $g=0$ in eq.~\eqref{eq:damruijgrok_u} yields
\bq
\label{eq:damruijgrok_u_g0}
[\dot{u}^\mu]_{g=0} = \frac{e}{m}\,F^{\mu\nu}u_\nu
-\frac{e}{m^2}(g^{\mu\nu} +u^\mu u^\nu) u^\alpha\partial_\alpha \tilde{F}_{\nu\beta} W^\beta \,.
\eq
Observe that the Stern-Gerlach force is nonzero 
even though the particle has no magnetic dipole moment.
The Stern-Gerlach force in this case arises entirely from the Thomas precession.

\eit
In the rest of this paper,
we shall employ the orbital Hamiltonian and spin precession vector in eq.~\eqref{eq:horb_spin},
and we do not set the speed of light to unity.

%\vfill\pagebreak
\setcounter{equation}{0}
\section{Passage through localized region with a field gradient}
\subsection{General remarks}
To make contact with the analysis in \cite{Talman_arxiv_Nov2016}, let the coordinate system be $(x,y,z)$, 
where a particle propagates at speed $v$ in the positive $x$ direction.
As in \cite{Talman_arxiv_Nov2016}, we employ a `hard edge' approximation
where the particle passes through a static magnetic or electric field which vanishes outside a region of length $L_q$.
The magnetic or electric field lies in the $(y,z)$ plane.
Treating only the Stern-Gerlach force, the equation of motion for $p_j$ is
\bq
\label{eq:dpdtloc}
\frac{dp_j^{SG}}{dt} = -\frac{\partial (\bm{\Omega}\cdot\bm{s})}{\partial r_j} \qquad (j=2,3) \,.
\eq
As in \cite{Talman_arxiv_Nov2016}, to the required level of approximation
the transverse momentum transfer is calculated via an impulse approximation.
The change to the $(y,z)$ coordinates and the spin is negligible (these are reasonable approximations).
The impulse $\Delta p_j^{SG}$ is given by multiplying the force by the time of flight $\Delta t = L_q/v$, viz.
\bq
\label{eq:kickpj}
\Delta p_j^{SG} \simeq \frac{dp_j}{dt}\,\Delta t 
= -\frac{\partial (\bm{\Omega}\cdot\bm{s})}{\partial r_j}\,\frac{L_q}{v} \qquad (j=2,3) \,.
\eq
We analyze the four cases treated in \cite{Talman_arxiv_Nov2016} in turn,
viz.~a magnetic `skew quadrupole' field,
a magnetic `erect quadrupole' field,
an electrostatic `erect quadrupole' field,
and finally an electrostatic `skew quadrupole' field.

\subsection{Magnetic skew quadrupole}
The vector potential is $\bm{A} = -k_byz\,\hat{\bm{x}}$ and the magnetic field is given by
\bq
\bm{B} = \bm{\nabla}\times\bm{A} = k_b(-y \hat{\bm{y}} +z\hat{\bm{z}}) \,.
\eq
The spin-orbit term in the Hamiltonian is, to the relevant level of approximation,
\bq
\begin{split}
\bm{\Omega}\cdot\bm{s}
&\simeq -\biggl(a +\frac{1}{\gamma}\biggr) \frac{e}{mc} \bm{B}\cdot\bm{s}
= -\biggl(a +\frac{1}{\gamma}\biggr) \frac{ek_b}{mc} (-ys_y +zs_z) \,.
\end{split}
\eq
From eq.~\eqref{eq:kickpj}, the impulses $\Delta p_y^{SG}$ and $\Delta p_z^{SG}$ are given by 
\begin{subequations}
\label{eq:mag_skew_me}
\begin{align}
\Delta p_y^{SG} &\simeq -\biggl(a +\frac{1}{\gamma}\biggr) \frac{e}{mc}\,s_yk_b\,\frac{L_q}{v} \,,
\\
\Delta p_z^{SG} &\simeq \phantom{-}\biggl(a +\frac{1}{\gamma}\biggr) \frac{e}{mc}\,s_zk_b\,\frac{L_q}{v} \,.
\end{align}
\end{subequations}
The corresponding expressions in \cite[eq.~(34)]{Talman_arxiv_Nov2016} are
\bq
\label{eq:mag_skew_talman}
\Delta p^2 = -\mu_e^* \hat{s}^{*2} k_b\,\frac{L}{v} \,,\qquad
\Delta p^3 = \mu_e^* \hat{s}^{*3} k_b\,\frac{L}{v} \,.
\eq
Note that the superscripts in \cite{Talman_arxiv_Nov2016}, hence in eq.~\eqref{eq:mag_skew_talman},
denote coordinate indices, not powers.
As stated in the introduction, the transverse momentum transfer (impulse) in \cite{Talman_arxiv_Nov2016} 
is proportional to the particle magnetic dipole moment and omits the contribution from the Thomas precession.
Expressing the above in terms of our notation,
\bq
\mu_e^* \hat{s}^{*2} = (a+1)\frac{e}{mc}\,s_y \,,\qquad
\mu_e^* \hat{s}^{*3} = (a+1)\frac{e}{mc}\,s_z \,.
\eq
In our notation, the expressions in eq.~\eqref{eq:mag_skew_talman} yield
\begin{subequations}
\label{eq:mag_skew_talman1}
\begin{align}
\Delta p^2 &= -(a_e+1) \frac{e}{mc}\,s_yk_b\,\frac{L_q}{v} \,,
\\
\Delta p^3 &= \phantom{-}(a_e+1) \frac{e}{mc}\,s_zk_b\,\frac{L_q}{v} \,.
\end{align}
\end{subequations}
These are to be compared with the expressions in eq.~\eqref{eq:mag_skew_me}.
The two sets of results coincide in the nonrelativistic limit $\gamma\to1$.

\subsection{Magnetic erect quadrupole}
The analysis is similar to that for a skew quadrupole.
The vector potential is now $\bm{A} = \frac12 k_b(y^2-z^2)\,\hat{\bm{x}}$ and the magnetic field is given by
\bq
\bm{B} = \bm{\nabla}\times\bm{A} = -k_b(z \hat{\bm{y}} +y\hat{\bm{z}}) \,.
\eq
The spin-orbit term in the Hamiltonian is, to the relevant level of approximation,
\bq
\begin{split}
\bm{\Omega}\cdot\bm{s}
&\simeq -\biggl(a +\frac{1}{\gamma}\biggr) \frac{e}{mc} \bm{B}\cdot\bm{s}
= \biggl(a +\frac{1}{\gamma}\biggr) \frac{ek_b}{mc} (zs_y +ys_z) \,.
\end{split}
\eq
From eq.~\eqref{eq:kickpj}, the impulses $\Delta p_y^{SG}$ and $\Delta p_z^{SG}$ are given by 
\begin{subequations}
\label{eq:mag_erect_me}
\begin{align}
\Delta p_y^{SG} &\simeq -\biggl(a +\frac{1}{\gamma}\biggr) \frac{e}{mc}\,s_zk_b\,\frac{L_q}{v} \,,
\\
\Delta p_z^{SG} &\simeq -\biggl(a +\frac{1}{\gamma}\biggr) \frac{e}{mc}\,s_yk_b\,\frac{L_q}{v} \,.
\end{align}
\end{subequations}
The corresponding expressions in \cite[eq.~(38)]{Talman_arxiv_Nov2016} are
\bq
\label{eq:mag_erect_talman}
\Delta p^2 = -\mu_e^* \hat{s}^{*3} k_b\,\frac{L}{v} \,,\qquad
\Delta p^3 = -\mu_e^* \hat{s}^{*2} k_b\,\frac{L}{v} \,.
\eq
These expressions are also proportional to the particle magnetic dipole moment and yield, in our notation,
\begin{subequations}
\label{eq:mag_erect_talman1}
\begin{align}
\Delta p^2 &= -(a_e+1) \frac{e}{mc}\,s_zk_b\,\frac{L_q}{v} \,,
\\
\Delta p^3 &= -(a_e+1) \frac{e}{mc}\,s_yk_b\,\frac{L_q}{v} \,.
\end{align}
\end{subequations}
These are to be compared with the expressions in eq.~\eqref{eq:mag_erect_me}.
As with a magnetic skew quadrupole, the two sets of results coincide in the nonrelativistic limit $\gamma\to1$.

\subsection{Electrostatic erect quadrupole}
The scalar potential is $\Phi = \frac12 k_e(y^2-z^2)$ and the electric field is given by
\bq
\bm{E} = -\bm{\nabla}\Phi = -k_e(y \hat{\bm{y}} -z\hat{\bm{z}}) \,.
\eq
The spin-orbit term in the Hamiltonian is, to the relevant level of approximation,
\bq
\begin{split}
\bm{\Omega}\cdot\bm{s}
&\simeq \biggl(a +\frac{1}{\gamma+1}\biggr) \frac{e}{mc} (\bm{\beta}\times\bm{E})\cdot\bm{s}
= -\biggl(a +\frac{1}{\gamma+1}\biggr) \frac{ek_ev}{mc^2} (zs_y +ys_z) \,.
\end{split}
\eq
From eq.~\eqref{eq:dpdtloc}, the equations for $dp_y^{SG}/dt$ and $dp_z^{SG}/dt$ are given by 
\begin{subequations}
\label{eq:elec_erect_me}
\begin{align}
\frac{dp_y^{SG}}{dt} &\simeq \biggl(a +\frac{1}{\gamma+1}\biggr) \frac{ek_ev}{mc^2}\,s_z \,,
\\
\frac{dp_z^{SG}}{dt} &\simeq \biggl(a +\frac{1}{\gamma+1}\biggr) \frac{ek_ev}{mc^2}\,s_y \,.
\end{align}
\end{subequations}
The corresponding expressions in \cite[eq.~(41)]{Talman_arxiv_Nov2016} are
(with an allowance for a factor of $c$ because we employ Gaussian units)
\bq
\label{eq:elec_erect_talman}
\frac{dp^2}{dt} = \frac{\mu_e^*k_ev}{c}\,\hat{s}^{*3} 
\,,\qquad
\frac{dp^3}{dt} = \frac{\mu_e^*k_ev}{c}\,\hat{s}^{*2} \,.
\eq
Expressing these in terms of our notation yields
\begin{subequations}
\label{eq:elec_erect_talman1}
\begin{align}
\frac{dp^2}{dt} &= (a_e+1)\frac{ek_ev}{mc^2}\,s_z \,,
\\
\frac{dp^3}{dt} &= (a_e+1)\frac{ek_ev}{mc^2}\,s_y \,.
\end{align}
\end{subequations}
These are to be compared with the expressions in eq.~\eqref{eq:elec_erect_me}.
In contrast to the case of magnetic quadrupoles, the two sets of results in
eqs.~\eqref{eq:elec_erect_me} and \eqref{eq:elec_erect_talman1}
do {\em not} coincide in the nonrelativistic limit $\gamma\to1$.
Neglecting the value of $a$, eq.~\eqref{eq:elec_erect_me} yields
\bq
\label{eq:elec_erect_me_nonrel}
\biggl[\frac{dp_y^{SG}}{dt}\biggr]_{\rm non-rel} \simeq \frac12\, \frac{ek_ev}{mc^2}\,s_z 
\,,\qquad
\biggl[\frac{dp_z^{SG}}{dt}\biggr]_{\rm non-rel} \simeq \frac12\, \frac{ek_ev}{mc^2}\,s_y \,.
\eq
However, eq.~\eqref{eq:elec_erect_talman1} yields
\bq
\label{eq:elec_erect_talman1_nonrel}
\biggl[\frac{dp^2}{dt}\biggr]_{\rm non-rel} = \frac{ek_ev}{mc^2}\,s_z 
\,, \qquad
\biggl[\frac{dp^3}{dt}\biggr]_{\rm non-rel} = \frac{ek_ev}{mc^2}\,s_y \,.
\eq
There is a difference of a factor of $\frac12$ between the two sets of formulas.
The factor of $\frac12$ is a well-known consequence of Thomas precession 
and was derived by Thomas himself: see \cite[eq.~(6.1)]{Thomas}
(which is for the case of an electron moving in a Coulomb electrostatic field).

\subsection{Electrostatic skew quadrupole}
The scalar potential is now $\Phi = k_eyz$ and the electric field is given by
\bq
\bm{E} = -\bm{\nabla}\Phi = -k_e(z \hat{\bm{y}} +y\hat{\bm{z}}) \,.
\eq
The spin-orbit term in the Hamiltonian is, to the relevant level of approximation,
\bq
\begin{split}
\bm{\Omega}\cdot\bm{s}
&\simeq \biggl(a +\frac{1}{\gamma+1}\biggr) \frac{e}{mc} (\bm{\beta}\times\bm{E})\cdot\bm{s}
= -\biggl(a +\frac{1}{\gamma+1}\biggr) \frac{ek_ev}{mc^2} (-ys_y +zs_z) \,.
\end{split}
\eq
From eq.~\eqref{eq:dpdtloc}, the equations for $dp_y^{SG}/dt$ and $dp_z^{SG}/dt$ are given by 
\begin{subequations}
\label{eq:elec_skew_me}
\begin{align}
\frac{dp_y^{SG}}{dt} &\simeq -\biggl(a +\frac{1}{\gamma+1}\biggr) \frac{ek_ev}{mc^2}\,s_y \,,
\\
\frac{dp_z^{SG}}{dt} &\simeq \phantom{-}\biggl(a +\frac{1}{\gamma+1}\biggr) \frac{ek_ev}{mc^2}\,s_z \,.
\end{align}
\end{subequations}
The corresponding expressions in \cite[eq.~(43)]{Talman_arxiv_Nov2016} are
(again with an allowance for a factor of $c$ because we employ Gaussian units)
\bq
\label{eq:elec_skew_talman}
\frac{dp^2}{dt} = -\frac{\mu_e^*k_ev}{c}\,\hat{s}^{*2} 
\,,\qquad
\frac{dp^3}{dt} = \frac{\mu_e^*k_ev}{c}\,\hat{s}^{*3} \,.
\eq
Expressing these in terms of our notation yields
\begin{subequations}
\label{eq:elec_skew_talman1}
\begin{align}
\frac{dp^2}{dt} &= -(a_e+1)\frac{ek_ev}{mc^2}\,s_y \,,
\\
\frac{dp^3}{dt} &= \phantom{-} (a_e+1)\frac{ek_ev}{mc^2}\,s_z \,.
\end{align}
\end{subequations}
These are to be compared with the expressions in eq.~\eqref{eq:elec_skew_me}.
As with an electrostatic erect quadrupole, the two sets of results in
eqs.~\eqref{eq:elec_skew_me} and \eqref{eq:elec_skew_talman1}
do not coincide in the nonrelativistic limit $\gamma\to1$.
Again neglecting the value of $a$,
the same factor of $\frac12$ which appeared for an electrostatic erect quadrupole also appears here.

%\vfill\pagebreak
\setcounter{equation}{0}
\section{\label{sec:centralpot} Spin-Orbit Coupling in a Central Potential}
In this section, we treat bounded nonrelativistic motion in a central potential.
We employ polar coordinates and denote the radial coordinate by $r$.
To make contact with the analysis in \cite{Talman_arxiv_Nov2016},
we treat the source of the potential as infinitely massive and neglect reduced mass effects.
The Hamiltonian is then (`$c$' for `central')
\bq
H_c = \frac{\bm{p}^2}{2m} + V(r) +\bm{\Omega}_c\cdot\bm{S} \,.
\eq
Here $V(r)$ can be any (attractive) central potential.
The spin operator will be denoted by $\bm{S}$.
The orbital angular momentum is $\bm{L} = \bm{r}\times\bm{p}$ and
the total angular momentum is $\bm{J}=\bm{L}+\bm{S}$.
There is no external magnetic field, hence in this model the spin precession vector is
\bq
\bm{\Omega}_c = -\frac{e}{mc} \biggl(a+\frac{1}{\gamma+1}\biggr)
\bm{E}\times\bm{\beta} \,.
\eq
The electric field for a central potential is radial and is given by
\bq
\bm{E} = -\frac{dV}{dr}\,\frac{\bm{r}}{r} \,.
\eq
The velocity is given by $\bm{v}=\bm{p}/m$.
For nonrelativistic motion, we set $\gamma=1$ so the spin-orbit coupling is 
\bq
\begin{split}
\bm{\Omega}\cdot\bm{S} 
&= \Bigl(a+\frac12\Bigr) \frac{e}{(mc)^2} \frac{1}{r}\frac{dV}{dr}\, (\bm{r}\times\bm{p})\cdot\bm{S}
\\
&= \Bigl(a+\frac12\Bigr) \frac{e}{(mc)^2} \frac{1}{r}\frac{dV}{dr}\, \bm{L}\cdot\bm{S} 
\\
&= \Bigl(a+\frac12\Bigr) \frac{e}{2(mc)^2} \frac{1}{r}\frac{dV}{dr}\, (\bm{J}^2 -\bm{L}^2 -\bm{S}^2) \,.
\end{split}
\eq
For a particle with spin $\frac12$, the value of $\bm{S}^2=\frac34\hbar^2$ is a constant.
Also $\bm{L}$ is a dynamical invariant for a rotationally invariant system, hence it commutes with the Hamiltonian.
For nonrelativistic motion in any central potential, the spin-orbit eigenstates are indexed by 
the eigenvalues of $\bm{J}^2$ and $\bm{L}^2$,
i.e.~$j(j+1)\hbar^2$ and $l(l+1)\hbar^2$, employing standard textbook notation.
We also know that for spin $\frac12$, $j=l\pm\frac12$
(except if $l=0$, then $j=\frac12$ only).

\bit
\item
For an orbital $S$ state, then $l=0$ by definition, hence $\bm{J}=\bm{S}$ and
$\bm{L}\cdot\bm{S}=\bm{J}^2-\bm{S}^2=0$.
There is no energy shift, or change to the eigenstates, for an orbital $S$ state.
In particular, the ground state of the hydrogen atom is an $S$ state, and the energy shift for this state is zero.
For a hydrogen atom, $V(r)$ is a Coulomb potential.

\item
However, it is stated in 
\cite[text before eq.~(44)]{Talman_arxiv_Nov2016}
``For simplest comparison we will consider a hydrogen atom
$Z=1$, in the lowest Bohr model semi-classical case, having
$n = 1$.''
A nonzero energy shift is then derived for the ground state of a hydrogen atom
\cite[eq.~(58)]{Talman_arxiv_Nov2016}
\bq
\frac{\Delta\mathcal{E}_1}{|\mathcal{E}_1|} = \pm \frac{\alpha^2}{6} \,.
\eq
Here $\alpha \simeq 1/137$ is the electromagnetic fine-structure constant,
and the $\pm$ signs are for the two spin orientations of the electron.
The ground state is split into a doublet.

\item
Formulas are then cited in \cite{Talman_arxiv_Nov2016} 
from a text by Leighton 
(see \cite[ref.~9]{Talman_arxiv_Nov2016} and \cite{Leighton1959} below)
``A somewhat similar, up-to-date, quantum mechanical
spin-orbit doublet separation calculation can be copied
from Leighton[9], assuming $Z=1$, $n = 1$, $l = 1$.''
See \cite[eqs.~(59-60)]{Talman_arxiv_Nov2016}.

\item
For the $n^{th}$ energy level of a hydrogen atom, the allowed values of $l$ are $l=0,1,\dots,n-1$.
Hence for the ground state $n=1$, the only allowed eigenvalue for the orbital angular momentum is $l=0$.
There is no Coulombic eigenstate with the quantum numbers $(n=1,l=1)$.
Leighton's text has been misquoted.

\eit
For completeness, let us treat briefly the case of a Coulomb potential $V = -Ze^2/r$.
Then
\bq
\frac{1}{r}\frac{dV}{dr} = \frac{Ze^2}{r^3} \,.
\eq
The energy shift of a Coulombic eigenstate is then given by
\bq
\begin{split}
\Delta \mathcal{E} &= \langle \bm{\Omega}\cdot\bm{S}\rangle
\\
& \propto (\bm{J}^2 -\bm{L}^2 -\bm{S}^2)
\int_0^\infty \frac{1}{r}\frac{dV}{dr} |\psi(r)|^2 \,r^2 dr
\\
&\propto 
(\bm{J}^2 -\bm{L}^2 -\bm{S}^2) \int_0^\infty \frac{|\psi(r)|^2}{r}\,dr \,.
\end{split}
\eq
Here $\psi(r)$ is the radial wavefunction.
This integral has been calculated in textbooks for the Coulombic eigenfunctions,
for example the text by Leighton \cite{Leighton1959}.
We saw that the energy shift vanishes for an orbital $S$ state,
hence the factor $1/r$ (as $r\to0$) does not pose a problem for these states.
For $l>0$, the radial wavefunction is proportional to powers of $r$ (associated Laguerre polynomials)
and the integral (i.e.~energy shift) is finite.
If $j=l+\frac12$ then $\bm{L}\cdot\bm{S} = l\hbar^2$
and if $j=l-\frac12$ (for $l>0$ only) then $\bm{L}\cdot\bm{S} = -(l+1)\hbar^2$.

%\vfill\pagebreak
\setcounter{equation}{0}
\section{\label{sec:conc} Conclusion}
The spin-orbit coupling term $\bm{\Omega}\cdot\bm{s}$ has been extensively validated in atomic and particle physics experiments.
Hamilton's equations then uniquely determine the orbital and spin motion, including the Stern-Gerlach force.
Covariant equations of motion for the orbit and spin were
derived in \cite{DamRuijgrok1980},
and the equations therein also demonstrate that
the Stern-Gerlach force is not proportional
to the particle's magnetic dipole moment.
The Hamiltonian formalism was applied to derive expressions for the transverse impulse
for passage through a region with a localized static magnetic or electric field gradient.
The corresponding analysis in \cite{Talman_arxiv_Nov2016}
neglects the contribution of the Thomas precession.
A numerical example is treated in \cite[eq.~(61)]{Talman_arxiv_Nov2016}, 
for a beam of 6 MeV electrons passing through a beamline of magnetic quadrupoles.
For 6 MeV electrons, $\gamma \simeq 12$ and $a_e \simeq 0.001159 \ll 1/\gamma$.
The Stern-Gerlach angular deflection to the electron motion
is therefore approximately a factor of 12 smaller than the value estimated in \cite{Talman_arxiv_Nov2016}.

The problem of the spin-orbit coupling for bounded nonrelativistic motion in a central potential was also briefly studied.
It was noted that the energy shift is zero for an orbital $S$ state, where $l=0$ for the orbital angular momentum.
Contrary to the derivation in \cite{Talman_arxiv_Nov2016},
the spin-orbit coupling does not split the ground state of a hydrogen atom into a doublet.

%\vfill\pagebreak


\begin{thebibliography}{99}

\bibitem{MottMassey}
N.~F.~Mott and H.~S.~W.~Massey, {\em The Theory of Atomic Collisions} $3^{rd}$ ed., Oxford University Press, Oxford (1965).

\bibitem{Kessler}
J.~Kessler, {\em Polarized Electrons} $2^{nd}$ ed., Springer, Berlin (1985).

\bibitem{Thomas}
L.~H.~Thomas, 
%The Kinematics of an electron with an axis,
{\it Philos.~Mag.} {\bf 3} 1--21 (1927).

\bibitem{JacksonCE3} 
J.~D.~Jackson, {\em Classical Electrodynamics} $3^{rd}$ ed., Wiley, New York (1998).

\bibitem{Talman_arxiv_Nov2016}
Richard Talman, 
%`Relativistic Stern-Gerlach Deflection,'
arXiv:1611.03810 [physics.acc-ph].
All statements in the text above pertain to version 1 of the post, available at
\url{https://arxiv.org/abs/1611.03810v1}

\bibitem{DamRuijgrok1980}
H.~Van Dam and Th.~W.~Ruijgrok,
%Classical relativistic equations for particles with spin moving in external fields,
{\it Physica A} {\bf 104} 281--297 (1980).

\bibitem{BMT}
V.~Bargmann, L.~Michel and V.~L.~Telegdi, 
%Precession of the polarization of particles moving in a homogeneous electromagnetic field,
{\it Phys.~Rev.~Lett.} {\bf 2} 435--436 (1959).

\bibitem{Leighton1959}
R.~B.~Leighton, {\em Principles of Modern Physics}, McGraw-Hill, New York, (1959).

\end{thebibliography}
\end{document}